\documentclass[12pt,leqno]{article}
\usepackage{amsmath}
\usepackage{amssymb,amscd,amsthm,amsfonts,amsgen,amsxtra,euscript,latexsym}
\usepackage{graphicx}
\usepackage{tikz-cd}

\usepackage{epsfig}   

\def\EE{H_{\rm evol}}

\newcommand{\ms}{\medskip}

\newcommand{\ba}{\begin{array}}
\newcommand{\ea}{\end{array}}

           \def\htop{{\text h_{{\rm top}}}}
 \def\Ptop{{\rm P_{\rm top}}}

\newcounter{bean}
    {
      \begin{list}{\bf #1(\arabic{bean})}
         {\usecounter{bean}
              \setcounter{bean}{-1}
          \labelsep=1em
          \settowidth{\labelwidth}{#1\thebean:}
          \addtolength{\labelwidth}{1.1ex} 
          \leftmargin=\labelwidth 
          \addtolength{\leftmargin}{\labelsep} }

    }    {\end{list}}

\usetikzlibrary{arrows.meta}


\makeatletter
\def\iint{\DOTSI\protect\ints@\tw@}
\def\iiint{\DOTSI\protect\ints@\thr@@}
\def\iiiint{\DOTSI\protect\ints@{4}}
\def\idotsint{\DOTSI\protect\ints@\z@}

\def\intkern@{\mkern-6mu\mathchoice{\mkern-3mu}{}{}{}}
\let\DOTSI\relax
\let\ilimits@\displaylimits

\def\ints@#1{%
  \mkern-7mu\mathchoice{\mkern-2mu}{}{}{}%
  \mathop{\mkern7mu\mathchoice{\mkern2mu}{}{}{}%
    \intop\ifnum#1=\z@\intdots@
    \else\intkern@\fi
    \ifnum#1>\tw@\intop\intkern@\fi
    \ifnum#1>\thr@@\intop\intkern@\fi
    \intop
  }\ilimits@
}
\makeatother

\newcommand{\cM}{\EuScript{M}}

\newcommand{\bR}{{\mathbb R}}

\newcommand{\cL}{{\mathcal L}}
\newcommand{\cR}{{\mathcal R}}

\newcommand{\Cross}{$\mathbin{\tikz [x=1.4ex,y=1.4ex,line width=.2ex] \draw (0,0) -- (0.5,0.5) (0,0.5) -- (0.5,0);}$}%

\makeatletter
\def\@maketitle{\newpage
 \null
 \vskip 2em
 \begin{center}%
  {\normalsize\bf \@title \par}%
  \vskip 1.5em
  {\normalsize
   \lineskip .5em
   \begin{tabular}[t]{c}\@author
   \end{tabular}\par}%
  \vskip 2em
  {\@date}%
 \end{center}%
 \par
 \vskip 2.5em}
\makeatother

\makeatletter
\renewcommand\section{\@startsection {section}{1}{\z@}%
                                   {-3.5ex \@plus -1ex \@minus -.2ex}%
                                   {2.3ex \@plus.2ex}%
                                   {\normalfont\normalsize\bfseries}}
\renewcommand\subsection{\@startsection{subsection}{2}{\z@}%
                                     {-3.25ex\@plus -1ex \@minus -.2ex}%
                                     {1.5ex \@plus .2ex}%
                                     {\normalfont\normalsize\bfseries}}

\makeatother

\begin{document}
\thispagestyle{empty}

\vspace*{3cm}

\begin{center}
\large{\bf Evolutionary entropy and the Second Law of Thermodynamics}

\vspace{1cm}
\begin{tabular}{lcl}
\textbf{ Lloyd A. Demetrius}&\quad  &\textbf{ Christian Wolf}\\
 & & \\
  Dept. of Organismic and& &Department of \\ 
 Evolutionary Biology& &Mathematics \\
   Harvard University& &The City Colleges of New York\\ 
  Cambridge, Mass. 02138, U.S.A.& &New York,  10031. U.S.A.
\end{tabular} 

\ms

May 20, 2020
  \end{center}

\newpage
\begin{abstract}
The  dynamics of molecular collisions in a macroscopic body are encoded by the parameter
 \textit{Thermodynamic entropy} --- a statistical measure of the number of molecular configurations that correspond  to a given macrostate.   Directionality in the flow of energy in macroscopic bodies is described by the Second Law of Thermodynamics: In isolated systems, that is systems closed to the input of energy and matter, thermodynamic entropy increases. 

\ms
The  dynamics of the lower level interactions  in populations of replicating  organisms is encoded by the parameter \textit{Evolutionary entropy}, a statistical measure which describes the number and diversity  of metabolic cycles in a population of replicating organisms. Directionality in the transformation of energy in populations of organisms is described by the Fundamental Theorem of Evolution: 
In systems open to the input of energy and matter, Evolutionary entropy increases, when the energy source is scarce and diverse, and decreases when the energy source is abundant and singular.

\ms
This article shows that when  $\rho \to 0$, and $N \to \infty$, where $\rho$ is the production rate of the external energy source, and $N$ denote the number of replicating units, evolutionary entropy, an organized state of energy; and thermodynamic entropy, a randomized state of energy, coincide. Accordingly, the Fundamental Theorem of Evolution, is a generalization  of the Second Law of Thermodynamics.
\end{abstract}

\newpage
\section{Introduction}

Most changes in the macroscopic properties of physical systems --- the universe of solids, liquids and gases, and the macroscopic behavior of biological systems --- the  world of macromolecules and cells, seem to be highly irregular, chaotic and unpredictable. There exist however, processes in both physical and biological systems which manifest predictable irreversible changes.

\ms
These changes are characterized by the flow and transformation of energy. In physical systems, the irreversibility is generated by the dynamics of molecular collisions in aggregates of inanimate matter. This process is encoded by the parameter, thermodynamic entropy, a statistical measure of the number of molecular configurations that correspond to a macroscopic state. Directionality in the flow of energy is described by the Second Law of Thermodynamics: Thermodynamic entropy increases in isolated systems, that is, systems that are closed to the input of energy and matter \cite{At}, \cite{Be}.

\ms

 Irreversibility in biological systems is depicted in terms of the transformation of energy from an external source to a population of replicating organisms. The transformation is driven by a variation-selection process which is encoded by the statistical parameter, evolutionary entropy, a measure of the number and diversity of metabolic cycles in a population of replicating organisms \cite{D2}, \cite{D3}. Directionality in this process of energy transformation is expressed by the Fundamental Theorem of Evolution: Evolutionary entropy increases when the energy source is limited and diverse, and decreases when the energy source is abundant and singular. 
 
 \ms
 
 Thermodynamic  entropy and evolutionary entropy are statistical measures of microscopic organization in aggregates of inanimate matter, and populations of replicating organisms, 
 respectively.
 
 \ms
 
 Thermodynamic entropy describes the extent to which energy is spread and shared among the molecules that comprise the macroscopic body. A solid has small entropy since the molecules occupy fixed positions within the macroscopic body. A gas has large entropy since the molecules are free to move around. Thermodynamic entropy can be considered a measure of positional disorder, a randomized state of energy.
 
 \ms
 Evolutionary entropy describes the  multiplicity of cyclic pathways of energy flow between the different metabolic states in a population of replicating organisms. A semelparous population (annual plants) consists of organisms that reproduce at a single stage in their life cycle.  Semelparous populations have entropy zero.
The energy transformation from birth to reproduction proceeds by  a single pathway. An iteroparous  population (perennial plants) consists of organisms that can reproduce at several distinct stages in their life-cycle. Energy transformation in perennials is described by a multiplicity of reproductive stages. An iteroparous population has positive entropy. Evolutionary entropy can be considered a measure of temporal order, and organized state of energy.
 
 \ms
 
 In this article we will describe an analytic relation between these two statistical measures of positional disorder, and temporal organization, respectively. Our analysis is based on the observation that a population of replicating organisms is an open system which maintains its viability by appropriating energy from an external energy source and transforming this chemical energy into biomass. We will show that when $\rho\to 0$ and $N\to \infty$, where $\rho$ is the production rate of the energy source, and $N$  the population size, then the Evolutionary entropy and Thermodynamic entropy coincide.
 
 \ms
 
 We will appeal to the relation between Thermodynamic entropy and Evolutionary entropy to show that the Fundamental Theorem of evolution is a generalization of the Second Law of Thermodynamics.
\ms 

The relation between the Second Law of Thermodynamics and the processes which underlie  the development and the evolution of living organisms, has been addressed by scientists from diverse disciplines, e.g., \cite{E}, \cite{Fi}, \cite{PNB}, \cite{Sch},  \cite{Wi}.

Prominent investigations in the Physical and Chemical Sciences include Schr\"odinger \cite{Sch}, who introduced the term negative entropy to describe the process whereby organisms maintain their ordered states by appropriating chemical energy from the environment, and Prigogine \cite{PNB}, who proposed the notion of entropy production to characterize the mechanism biological systems exploit to maintain coherent behavior of steady states far from thermodynamic equilibrium.

Investigators from Evolutionary genetics, have addressed the relation between the Second Law and the evolution of living organisms by proposing various measures of Darwinian fitness as analogues of the Boltzmann entropy. The proposition of the Malthusian parameter as a measure of Darwinian fitness, see Fisher \cite{Fi}, is one of the most influential efforts to relate the Second Law with   evolutionary dynamics. In Section 6 we reevaluate these investigations within the framework of evolutionary entropy and its relation with the Second Law.

\ms
This article is organized as follows: Section 2 provides a succinct account of the process which underlies energy flow in aggregates of inanimate matter. Section 3 gives an account of  energy transformation in populations of  organisms, and the origin of the concept Evolutionary entropy.
Section 4 presents the analysis  of a  mathematical model which forms the framework for the Fundamental Theorem of evolution. Finally, the relation between Thermodynamic entropy and Evolutionary entropy is analyzed in Section 5. This relation provides the analytical basis for the claim (Section 6) that the Second Law of Thermodynamics is the limiting case of the Fundamental Theorem of Evolution. 

\section{Energy flow in Physical Systems}

The main example of directionality in physical systems is the flow of heat between bodies at different temperatures. Empirical studies and experimental observations led Clausius to the proposition that the flow of energy as heat from a body at higher temperature to one at a lower temperature is a universal phenomenon. The flow is spontaneous in the sense that the process does not require the action of some kind of work, mechanical or electrical. The realization that the reverse process, namely the transfer of heat from a cold body to a hotter body is not spontaneous, constitutes an asymmetry in Nature.

\subsection{Thermodynamic Entropy}
Clausius appealed to mathematical arguments to show that this asymmetry entails the existence of a  property of matter  which he  called \textit{entropy.} The Clausius entropy admits an analytic description, namely
\begin{equation}\label{eq1}
dS_c =  \frac{dQ}{T}
\end{equation}
Here, $dQ$ is the small amount of heat added to a body with temperature $T$. The quantity $dS_c$ represents the increase in entropy.

\ms
Material aggregates such as a gas are composed of molecules which move according to the laws of classical mechanics. The issue of reconciling the time symmetric description of molecular dynamics with the time asymmetric description of the evolution of a  macroscopic system emerged as a fundamental problem as soon as Clausius' discovery was recognized. The problem was ultimately resolved by Boltzmann. 

\ms
The model proposed by Boltzmann was based on  the microscopic representation of the Clausius entropy. Boltzmann's analysis was based on the quantity 
\begin{equation}\label{eq2}
S= k_b\log W.
\end{equation}
Here $W$ denotes the number of microscopic configurations compatible with a given macrostate, and $k_b$, a constant, called Boltzmann's constant.

\ms
If $N$ denotes the number of molecules in the macroscopic body, then the number of configurations associated with a given macroscopic state is given by
\begin{equation}\label{eq3}
W=\left(\frac{N!}{n_1!n_2!n_3!\ldots}\right).
\end{equation}
This follows from the fact that $W$ is the number of ways of distributing the $N$ molecules so that $n_1$ are in state 1, $n_2$ molecules in state 2 and so on.
\ms

The Boltzmann entropy of a solid and a gas are illustrated in Fig. 1 (a) and Fig. 1. (b), respectively.

\begin{center}
\begin{tikzpicture}

\draw[thick](0,0) rectangle (4,3);
\node at (1.8,-0.6) {Solid};
\node at (8.6,-0.6) {Gas};
\draw[thick](6.5,0) rectangle (10.5,3);
\filldraw[blue] (0.5,0.5) circle (0.075); \filldraw[blue]  (1,0.5) circle (0.075); \filldraw[blue]  (1.5,0.5) circle (0.075); \filldraw[blue]  (2,0.5) circle (0.075);\filldraw[blue]  (2.5,0.5) circle (0.075);\filldraw[blue]  (3,0.5) circle (0.075);\filldraw[blue]  (3.5,0.5) circle (0.075);
\filldraw[blue]  (0.5,1) circle (0.075); \filldraw[blue]  (1,1) circle (0.075); \filldraw[blue]  (1.5,1) circle (0.075); \filldraw[blue]  (2,1) circle (0.075);\filldraw[blue]  (2.5,1) circle (0.075);\filldraw[blue]  (3,1) circle (0.075);\filldraw[blue]  (3.5,1) circle (0.075);
\filldraw[blue]  (0.5,1.5) circle (0.075); \filldraw[blue]  (1,1.5) circle (0.075); \filldraw[blue]  (1.5,1.5) circle (0.075); \filldraw[blue]  (2,1.5) circle (0.075);\filldraw[blue]  (2.5,1.5) circle (0.075);\filldraw[blue]  (3,1.5) circle (0.075);\filldraw[blue]  (3.5,1.5) circle (0.075);
\filldraw[blue]  (0.5,2) circle (0.075); \filldraw[blue]  (1,2) circle (0.075); \filldraw[blue]  (1.5,2) circle (0.075); \filldraw[blue]  (2,2) circle (0.075);\filldraw[blue]  (2.5,2) circle (0.075);\filldraw[blue]  (3,2) circle (0.075);\filldraw[blue]  (3.5,2) circle (0.075);
\filldraw[blue]  (0.5,2.5) circle (0.075); \filldraw[blue]  (1,2.5) circle (0.075); \filldraw[blue]  (1.5,2.5) circle (0.075); \filldraw[blue]  (2,2.5) circle (0.075);\filldraw[blue]  (2.5,2.5) circle (0.075);\filldraw[blue]  (3,2.5) circle (0.075);\filldraw[blue]  (3.5,2.5) circle (0.075);

\draw [thick,->](6.9,1.2) -- (6.9,1.6);
\draw [thick,->](7.9,0.4) -- (7.6,0.9);
\draw [thick,->](7,0.5) -- (7.3,0.8);
\draw [thick,->](8,1.5) -- (7.7,1.2);
\draw [thick,->](9,2.3) -- (9.5,2);
\draw [thick,->](8.8,2.6) -- (9.2,2.7);
\draw [thick,->](8.5,0.5) -- (9,0.6);
\draw [thick,->](10,0.5) -- (9.8,1.2);
\draw [thick,->](9.4,1.1) -- (9.7,0.8);
\draw [thick,->](9,1.5) -- (9,1.1);
\draw [thick,->](10.2,1.5) -- (9.6,1.5);
\draw [thick,->](7,0.5) -- (7.3,0.8);
\draw [thick,->](8,1.9) -- (7.6,1.9);
\draw [thick,->](7.2,2.7) -- (7.1,2.3);
\draw [thick,->](8,2.5) -- (7.4,2.4);
\draw [thick,->](8.3,1.5) -- (8.3,1.8);
\draw [thick,->](8.3,0.9) -- (8.6,1.3);
\draw [thick,->](8.7,2.2) -- (8.6,2.8); 
\draw [thick,->](9,1.8) -- (8.8,1.6);
\draw [thick,->](9.9,2.7) -- (9.6,2.6);
\draw [thick,->](9.6,2.1) -- (9.9,2.3);
\draw [thick,->](9.9,2.7) -- (9.6,2.6);
\draw [thick,->](7.3,1.8) -- (7.6,1.6);
\draw [thick,->](9.4,0.3) -- (9.3,.6);
\draw [thick,->](8,2.3) -- (8.3,2.2);
\end{tikzpicture}
\end{center}

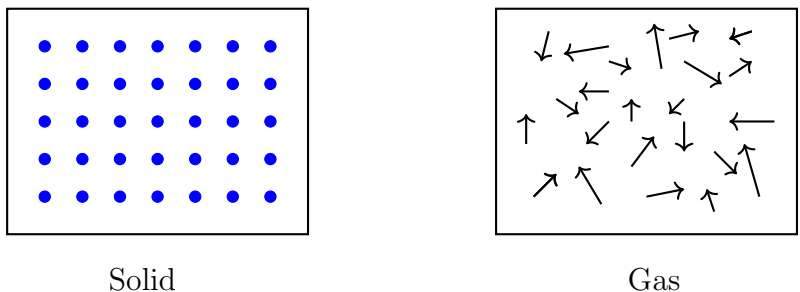
\centerline{ Fig. 1. (a), 1 (b) }

The molecules in a solid occupy a fixed position in the macroscopic body. The entropy S as computed in Eq. \eqref{eq2} is small. The molecules in the gas are free to move around, the entropy S  is large.

\subsection{Directionality in Physical Systems}
The evolutionary dynamics of the Boltzmann entropy was analyzed under the following assumptions:
\begin{enumerate}
	\item[(a)]  The molecules of the gas, which is assumed to be confined in a container, move randomly.
	\item[(b)] The velocities of the molecules are randomly distributed .
 \item[(c)] The molecules collide with each other and velocities change after collision.
	\end{enumerate}
Boltzmann showed that if the number of molecules in the system is immensely large, and the system evolves in an isolated environment, the quantity $S$ will increase to an equilibrium state. 

\ms
In systems which are isolated, that is closed to the input of energy and matter, the analysis of the dynamics of the interacting molecules shows that $S$ the degree of energy spreading and sharing among the microscopic storage modes of the system, satisfies the principle
\begin{equation}\label{eq4}
\Delta S \ge 0. 
\end{equation}
Since the Boltzmann entropy, $S$, is identical to the Clausius entropy defined in \eqref{eq1}, the relation \eqref{eq4} can be interpreted as a molecular dynamical explanation of the asymmetry of the flow of heat energy.

\ms

The evolution of an isolated macroscopic system evolving in time, is exemplified by the macroscopic diversity profile of a fluid in the three frames in Fig. 2. The dots in the figure represent the density variable of the fluid at different times during the evolution of the system. The evolution can be considered as the flow of heat energy. The left half of the system
in Fig. 2 (a) is hotter than the right half. In Fig. 2 (c) the temperature is uniform.

\ms

\begin{center}
\begin{tikzpicture}

\draw[thick](0,0) rectangle (3.8,2.7);
\draw[thick](1.9,0) -- (1.9,1.2);
\draw[thick](1.9,1.5) rectangle (1.9,2.7);

 \foreach \x in {1,...,170}
    {
      \pgfmathrandominteger{\a}{10}{180}
      \pgfmathrandominteger{\b}{10}{260}
      \fill (\a*0.01,\b*0.01) circle (0.03);
    };

\draw[thick](5.2,0) rectangle (9,2.7);
\draw[thick](7.1,0) -- (7.1,1.2);
\draw[thick](7.1,1.5) rectangle (7.1,2.7);

 \foreach \x in {1,...,150}
    {
      \pgfmathrandominteger{\a}{10}{180}
      \pgfmathrandominteger{\b}{10}{260}
      \fill (\a*0.01+5.2,\b*0.01) circle (0.03);
    };
    
     \foreach \x in {1,...,20}
    {
      \pgfmathrandominteger{\a}{10}{180}
      \pgfmathrandominteger{\b}{10}{260}
      \fill (\a*0.01+7.1,\b*0.01) circle (0.03);
    };

\draw[thick](10.6,0) rectangle (14.4,2.7);
\draw[thick](12.5,0) -- (12.5,1.2);
\draw[thick](12.5,1.5) rectangle (12.5,2.7);

 \foreach \x in {1,...,85}
    {
      \pgfmathrandominteger{\a}{10}{180}
      \pgfmathrandominteger{\b}{10}{260}
      \fill (\a*0.01+10.6,\b*0.01) circle (0.03);
    };
    
     \foreach \x in {1,...,85}
    {
      \pgfmathrandominteger{\a}{10}{180}
      \pgfmathrandominteger{\b}{10}{260}
      \fill (\a*0.01+12.5,\b*0.01) circle (0.03);
    };

\node at (1.9,-0.6) {(a)};
\node at (7.1,-0.6) {(b)};
\node at (12.5,-0.6) {(c)};

\end{tikzpicture}
\end{center}

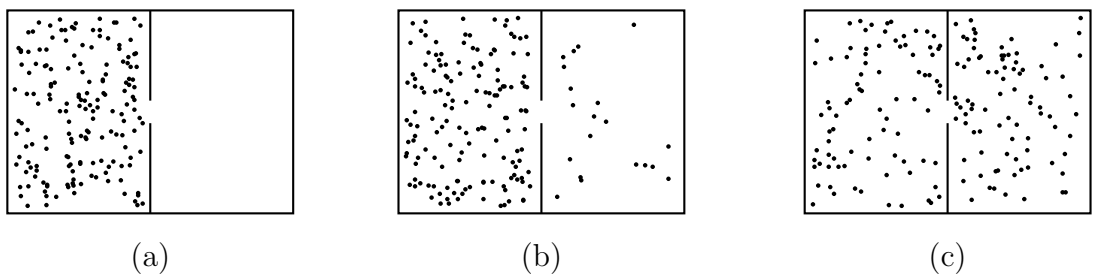
\centerline{ Fig. (2) The macroscopic density profile of an isolated system at three different times.}

\section{Energy transformation in Biological Systems}

\ms
Heat is the simplest and most frequently used medium by which energy is transformed in aggregates of inanimate matter: solids, liquid and gases.. However, there is a fundamental restriction which exists on the conversion of thermal energy into work. This restriction is expressed by the relation
\begin{equation}\label{eq5}
\omega=q\left(\frac{T_2-T_1}{T_1}\right).
\end{equation}
Here $\omega$ is the maximal work derived and $q$ is the absorbed heat, and $T_2$ and $T_1$ are the absolute temperatures of the material bodies between which the heat passes.

\ms
The relation \eqref{eq5} has significant implications in the study of energy transformations in biological systems: There is almost no temperature differential between the cells in a tissue or between the tissues in an organism. This implies that in living matter, thermal energy cannot be effectively transformed into work.

\ms
Living organisms are essentially isothermal chemical machines, Lehninger (1965) \cite{Leh}. The chemical components of these machines are not in thermodynamic equilibrium but in a dynamic steady state:

\ms 

The critical parameter in the dynamics of energy transformation in living organisms is not \underline{temperature}, the mean kinetic energy of the molecules in a macroscopic body, but \underline{cycle time}, the mean turn over time of the metabolic entities that comprise the population.

\subsection{Evolutionary Entropy}

Living organisms differ from material aggregates not only in terms of their isothermal character, but also in terms of their interaction with the external environment.

\ms
Organisms maintain their integrity by appropriating resources from the external environment and transforming this chemical energy into metabolic energy and biomass.
\ms
Organisms can be classified in terms of various states: age; size; metabolic energy. A population, the fundamental unit of the evolutionary process, can be represented as a
network or a directed graph, Fig. (3).

\begin{center}

\begin{tikzpicture}[scale=.75]
\draw[thick] (0,2.36) circle (0.25);
\node at (0,2.5) (1) {};
\draw[thick] (-3.5,0) circle (0.25);
\node at (-3,0.15) (B) {};

\draw[thick]  (0.2,-2) circle (0.25);
\node at (-2,0) (C) {};

\draw[thick]  (3.7,-0.3) circle (0.25);
\node at (-2,0) (D) {};

\draw[thick]  (4.3,3) circle (0.25);

\draw[thick] (-0.37,2.38) edge[->,in=80,out=160] (-3.5,0.35);
\draw[thick] (-3.13,0.06) edge[->,in=-130,out=20] (-0.15,2.03);

\draw[thick] (-3.5,-0.45) edge[->,in=-160,out=-75] (-0.15,-2.);
\draw[thick] (0.07,2.02) edge[->,in=85,out=-65] (0.2,-1.65);
\draw[thick] (0.13,2.05) edge[->,in=180,out=-65] (3.35,-0.32);
\draw[thick] (4,3.02) edge[->,in=25,out=165] (0.35,2.35);

\draw[thick] (0.56,-2.03) edge[->,in=230,out=0] (3.56,-0.59);

\draw[thick] (3.9,0) edge[->,in=-70,out=50] (4.4,2.7);

\draw[thick]  (4.1,2.7) edge[->,in=100,out=-120](3.65,0.02);

\draw[thick](1) edge  [->,in=120,out=60,loop,{looseness=45}]  (1);


\node[above] at (0,2.02) {\footnotesize{1}};
\node[left] at (-3.18,0.03) {\footnotesize{2}};
\node[below] at (0.2,-1.63) {\footnotesize{3}};
\node[right] at (3.39,-0.3) {\footnotesize{5}};
\node[right] at (3.98,3.02) {\footnotesize{4}};
\end{tikzpicture}

\end{center}

\ms
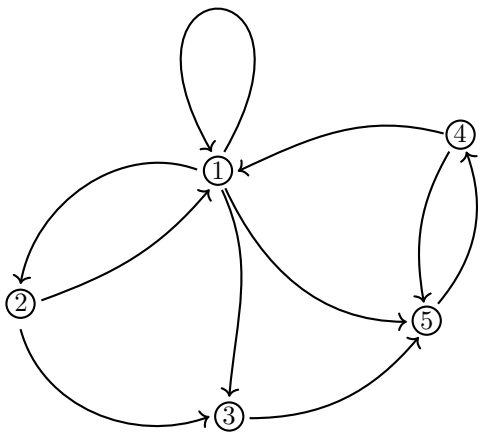
\centerline{Fig. (3) Population as a directed graph.}
\ms

\ms
The nodes of the graph correspond to the different states; the links between the nodes describe the interaction between the individuals that belong to the different states. The graph represents the transfer of energy between the individuals that define the different states.

\ms Evolutionary entropy, $H$,  a concept which has its mathematical roots and its biological rationale in the ergodic theory of dynamical systems \cite{D3}, can be expressed as 
\begin{equation}
H=\frac{\widetilde{S}}{\widetilde{T}}.
\end{equation}
The quantity $\widetilde{S}$ is called population entropy and $\widetilde{T}$ is the mean cycle time \cite{D2}. 
\ms
These quantities can be formally described by considering the set of nodes of the graph, denoted by $X=(1,2,\dots,d).$ We now fix an arbitrary vertex $a\in X$ and consider the set $X^\ast$ of all directed paths which start at $a$, end at $a$, and do not visit $a$ in the middle. An element $\tilde{a}$ in $X^\ast$ is written as  $a\to \beta_1 \to \beta_2\to \cdots \to \beta_{n-1}\to a$ which we denote by
\begin{equation}
\tilde{a}=[a \beta_1 \beta_2 \cdots  \beta_{n-1} a].
\end{equation}
We define $\phi_{\tilde{a}}=\phi_{a\beta_1}\phi_{\beta_1\beta_2}\cdots \phi_{\beta_{n-1}a}$. The quantities $\widetilde{S}$ and $\widetilde{T}$ are given by
\begin{equation}
\widetilde{S}=- \sum_{\tilde{a}\in X^\ast} \phi_{\tilde{a}} \log \phi_{\tilde{a}}
\end{equation}
and 
\begin{equation}
\widetilde{T}= \sum_{\tilde{a}\in X^\ast} \vert \tilde{a} \vert \phi_{\tilde{a}},
\end{equation}
where $\vert \tilde{a} \vert$ is the length of the path $\tilde{a}$.

	In the case of a population in which the individuals are divided into age--classes, the population entropy  $\widetilde{S}$ is given by
\begin{equation}\label{eq6}
	\widetilde{S}=-\sum_j p_j\log p_j
	\end{equation}
Here $p_j$ is the probability that the mother of  a randomly  chosen new born belongs to the age--class $(j)$.	 
\ms
The life-cycle of a population of annual plants and the population of perennial plants are described in Fig. 4(a) and Fig. 4(b) respectively. 
\begin{center}
\begin{tikzpicture}[scale=.6]
\node at (0,0) (A) {\footnotesize{1}};
\node at (2,0) (B) {\footnotesize{2}};
\node at (4,0) (C) {\footnotesize{3}};
\node at (5.2,0) (D) {};
\node at (6.8,0) (De) {};
\node at (8,0) (E) {\footnotesize{d}};
\node at (-1.1,1.95)(F){};
\draw[thick] (A) circle (.4);
\draw[thick] (B) circle (.4);
\draw[thick] (C) circle (.4);
\draw[thick] (E) circle (.4);

\draw[thick] (A) edge[->] node[below]{} (B);
\draw[thick] (B) edge[->] node[below]{} (C);
\draw[thick] (C) edge[->] node[below]{} (D);
\draw[thick] (De) edge[->] node[below]{} (E);

\draw[thick] (E) edge[->,in=75,out=105] node[above right]{} (A);

\draw[ultra thick, loosely dotted] (D)  --   (De);

\node at (3.7,-2.2) {Fig. 4(a)};

\node at (12,0) (A2) {\footnotesize{1}};
\node at (14,0) (B2) {\footnotesize{2}};
\node at (16,0) (C2) {\footnotesize{3}};
\node at (17.2,0) (D2) {};
\node at (18.8,0) (De2) {};
\node at (20,0) (E2) {\footnotesize{d}};

\draw[thick] (A2) circle (.4);
\draw[thick] (B2) circle (.4);
\draw[thick] (C2) circle (.4);
\draw[thick] (E2) circle (.4);

\draw[thick] (A2) edge[->] node[below]{} (B2);
\draw[thick] (B2) edge[->] node[below]{} (C2);
\draw[thick] (C2) edge[->] node[below]{} (D2);
\draw[thick] (De2) edge[->] node[below]{} (E2);
\draw[thick] (B2) edge[->,in=30,out=150] node[above right]{} (A2);
\draw[thick] (C2) edge[->,in=55,out=125] node[above right]{} (A2);
\draw[thick] (E2) edge[->,in=75,out=105] node[above right]{} (A2);

\draw[ultra thick, loosely dotted] (D2)  --   (De2);

\node at (15.7,-2.2) {Fig. 4(b)};

\end{tikzpicture}
\end{center}

The population entropy of annuals is $\widetilde{S}=0$. The energy flow in the population is described by a unique cycle. The perennials, as shown in Fig. 4(b), reproduce at several distinct stages in the life cycle. Energy transformation within the population is described by several distinct metabolic cycles.

\ms

The recurrence time in an age--structured  population is the generation time, the mean age of mothers at the birth of their offspring. This is given by
\begin{equation}\label{eq7}
\widetilde{T}=\sum_j jp_j
\end{equation}

\subsection{Directionality in Biological Systems}

\ms
Organisms are metabolic entities endowed with a genome which, together with the environment, regulates individual behavior. The flow of energy in a population of replicating organisms is modulated by adaptive, dynamical processes. These processes, which have no counterpart in physical systems, derive from the intrinsic instability of organic molecules, such as DNA, RNA and proteins, and competition between the organisms for an external energy source. These processes are:
\begin{enumerate}
	\item[(i)] \textit{Mutation: } Random changes in DNA; the genetic endowment of the organisms.\\
	The effect of mutation on the genome will be the emergence of a population consisting of two types of organisms: an ancestral type of large size, a variant of small size with the genetic endowment of the mutant.

\item[(ii)]\textit{Selection: } Competition between the ancestral type with entropy $H$ and the variant type with entropy $H^\ast$ will result in a change in the genetic and phenotypic composition of the population.

\end{enumerate} 
\ms 

The directional change, $\Delta H$ in the evolutionary entropy is described by the relation, \cite{D1}, \cite{D2}
\begin{equation}\label{eq9}
\left(-\Phi+ \frac{\gamma}{M}\right)\Delta H>0.
\end{equation}
The parameters $\Phi$ and $\gamma$, are macroscopic parameters; functions of the microlocal variables which describe the interaction between the organisms. These parameters are correlated with the resource endowment of the 
environment.

\ms
The parameter $\Phi$ corresponds to the resource amplitude:

\begin{enumerate}
\item[(i)] $\Phi < 0$: Scarcity; $\Phi > 0$, abundance. 
\end{enumerate}
The parameter $\gamma$ corresponds to the resource composition.
\begin{enumerate}
\item[(ii)]  $\gamma <0$: singular; $\gamma > 0$ diverse.
\end{enumerate}
The quantity $M$ denotes the population size, whereas $\Delta H = H^*-H$, $H$ and $H^*$ denotes the evolutionary entropy of the incumbent and the variant population respectively.

\ms

The relation \eqref{eq9} is a derivative of the  \underline{Entropic Selection Principle}: The outcome of competition between an incumbent population and a variant is contingent on the resource endowment and is
predicted by Evolutionary entropy.

\ms

The principle can be graphically exemplified by the profiles of a variant and and incumbent population. Fig. (5) and (6) describe the changes in profile due to the introduction of a mutant which ultimately replaces the
ancestral type. Fig. (5) describes the situation where the resource endowment is limited and diverse. Fig. (6) represents the condition where the resource endowment is abundant and singular. \\

\begin{center}
\begin{tikzpicture}

\draw[thick](0,0) rectangle (2.5,2.5);

\pgfmathsetmacro{\lc}{0.3}

 \foreach \y in {0,...,3}
 {

 \foreach \x in {0,...,3}
    {
      \pgfmathrandominteger{\a}{-1}{1}
      \pgfmathrandominteger{\b}{-1}{1}
        \pgfmathrandominteger{\r}{5}{7}

      \fill (\lc+\x*0.6+\a*0.06,0.3+\y*0.6+\b*0.06) circle (0.01*\r);
    };
};

\draw[thick](4.5,0) rectangle (7,2.5);

\pgfmathsetmacro{\lc}{4.8}

 \foreach \y in {0,...,3}
 {

 \foreach \x in {0,...,3}
    {
      \pgfmathrandominteger{\a}{-1}{1}
      \pgfmathrandominteger{\b}{-1}{1}
        \pgfmathrandominteger{\r}{5}{7}
            \pgfmathrandominteger{\po}{1}{25}
  \pgfmathparse{\po< 21 ? int(1) : int(0)}

 \ifnum\pgfmathresult=1 
           \fill (\lc+\x*0.6+\a*0.06,0.3+\y*0.6+\b*0.06) circle (0.01*\r);

    \else
          \node at (\lc+\x*0.6+\a*0.05,0.4+\y*0.6+\b*0.05){\Cross};

    \fi

    };
};

\draw[thick](9,0) rectangle (11.5,2.5);

\draw[thick](4.5,0) rectangle (7,2.5);

\pgfmathsetmacro{\lc}{9.2}

 \foreach \y in {0,...,3}
 {

 \foreach \x in {0,...,3}
    {
      \pgfmathrandominteger{\a}{-1}{1}
      \pgfmathrandominteger{\b}{-1}{1}
        \pgfmathrandominteger{\r}{5}{7}
            \pgfmathrandominteger{\po}{1}{25}
  \pgfmathparse{\po< 20 ? int(1) : int(0)}

 \ifnum\pgfmathresult=1 
   \node at (\lc+\x*0.6+\a*0.05,0.4+\y*0.6+\b*0.05){\Cross};
            \else
               \fill (\lc+\x*0.6+\a*0.06,0.3+\y*0.6+\b*0.06) circle (0.01*\r);

    \fi

    };
};

\draw[thick](13.5,0) rectangle (16,2.5);

\draw[thick,->] (3,1.25) -- (4,1.25);
\draw[thick,->] (7.5,1.25) -- (8.5,1.25);
\draw[thick,->] (12,1.25) -- (13,1.25);

\pgfmathsetmacro{\lc}{13.9}

 \foreach \y in {0,...,3}
 {

 \foreach \x in {0,...,3}
    {
      \pgfmathrandominteger{\a}{-2}{2}
      \pgfmathrandominteger{\b}{-2}{2}
        \pgfmathrandominteger{\r}{4}{7}

      \node at (\lc+\x*0.6+\a*0.05,0.4+\y*0.6+\b*0.05){\Cross};
    };
};

\node at (1.25,-0.6) {(a)};
\node at (5.75,-0.6) {(b)};
\node at (10.25,-0.6) {(c)};
\node at (14.75,-0.6) {(d)};

\end{tikzpicture}
\end{center}

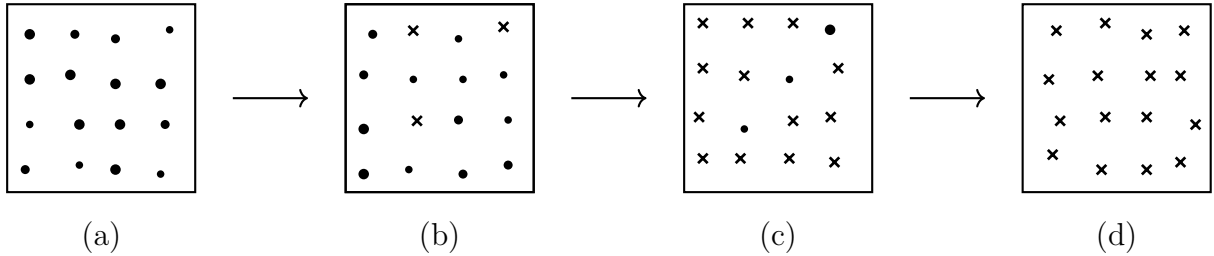
\centerline{ Fig. (5) Evolution under limited, diverse resources: $\Delta H>0$.}

\ms

Fig. 5(a): Ancestral type: Low entropy.\\
Fig. 5(b): Introduction of mutant  - higher entropy.\\
Fig. 5(c) Increase in frequency of mutant.\\
Fig. 5(d) Replacement of the ancestral type by variant.

\ms

\begin{center}
\begin{tikzpicture}

\draw[thick](0,0) rectangle (2.5,2.5);

\pgfmathsetmacro{\lc}{0.3}

 \foreach \y in {0,...,3}
 {

 \foreach \x in {0,...,3}
    {
    
      \pgfmathrandominteger{\a}{-2}{2}
      \pgfmathrandominteger{\b}{-2}{2}
        \pgfmathrandominteger{\r}{4}{7}
  \pgfmathrandominteger{\p}{1}{20}

         \node at (\lc+\x*0.6+\a*0.05,0.4+\y*0.6+\b*0.05){\Cross};

    };
};

\draw[thick](4.5,0) rectangle (7,2.5);

\pgfmathsetmacro{\lc}{4.7}

 \foreach \y in {0,...,3}
 {

 \foreach \x in {0,...,3}
    {
      \pgfmathrandominteger{\a}{-1}{1}
      \pgfmathrandominteger{\b}{-1}{1}
        \pgfmathrandominteger{\r}{5}{7}
            \pgfmathrandominteger{\po}{1}{25}
  \pgfmathparse{\po< 21 ? int(1) : int(0)}

 \ifnum\pgfmathresult=1 
         
 \node at (\lc+\x*0.6+\a*0.05,0.4+\y*0.6+\b*0.05){\Cross};
    \else
         
  \fill (\lc+\x*0.6+\a*0.06,0.3+\y*0.6+\b*0.06) circle (0.01*\r);
    \fi

    };
};

\draw[thick](9,0) rectangle (11.5,2.5);

\pgfmathsetmacro{\lc}{9.2}

 \foreach \y in {0,...,3}
 {

 \foreach \x in {0,...,3}
    {
      \pgfmathrandominteger{\a}{-1}{1}
      \pgfmathrandominteger{\b}{-1}{1}
        \pgfmathrandominteger{\r}{5}{7}
            \pgfmathrandominteger{\po}{1}{25}
  \pgfmathparse{\po< 21 ? int(1) : int(0)}

 \ifnum\pgfmathresult=1 
 
  \fill (\lc+\x*0.6+\a*0.06,0.3+\y*0.6+\b*0.06) circle (0.01*\r);


    \else
     \node at (\lc+\x*0.6+\a*0.05,0.4+\y*0.6+\b*0.05){\Cross};

    \fi

    };
};

\draw[thick](13.5,0) rectangle (16,2.5);

\draw[thick,->] (3,1.25) -- (4,1.25);
\draw[thick,->] (7.5,1.25) -- (8.5,1.25);
\draw[thick,->] (12,1.25) -- (13,1.25);

\pgfmathsetmacro{\lc}{13.9}

 \foreach \y in {0,...,3}
 {

 \foreach \x in {0,...,3}
    {
    
      \pgfmathrandominteger{\a}{-1}{1}
      \pgfmathrandominteger{\b}{-1}{1}
        \pgfmathrandominteger{\r}{5}{7}

        \fill (\lc+\x*0.6+\a*0.06,0.3+\y*0.6+\b*0.06) circle (0.01*\r);
    };
};

\node at (1.25,-0.6) {(a)};
\node at (5.75,-0.6) {(b)};
\node at (10.25,-0.6) {(c)};
\node at (14.75,-0.6) {(d)};

\end{tikzpicture}
\end{center}
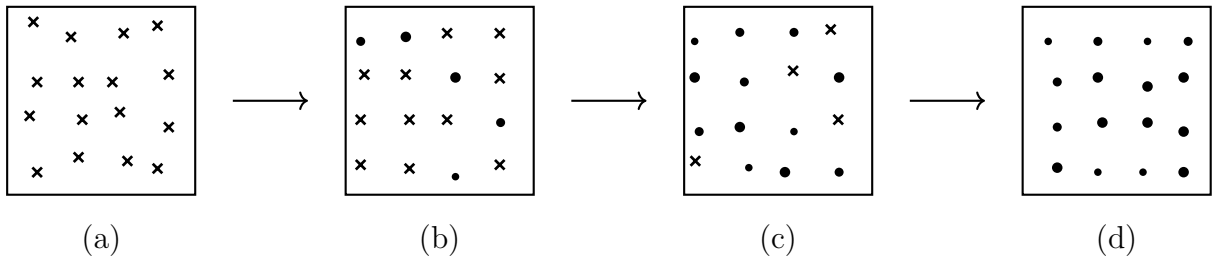
\centerline{ Fig. 6: Evolution under abundant, singular resource constraints: $\Delta H<0$.}
\ms

Fig. 6(a): Ancestral type: High entropy.\\
Fig. 6(b): Emergence of mutant  with low entropy.\\
Fig. 6(c) Increase in frequency of variant type.\\
Fig. 6(d) Replacement of the ancestral type by variant. 

\ms
The Entropic selection principle pertains to a local event- the competition between an incumbent and a variant population for the resources provided by the external environment.
\ms
The Fundamental Theorem of evolution describes the directional chances in evolutionary entropy as new variants are continuously introduced in the population and compete with the resident. This Theorem, refers to the long term changes in Evolutionary entropy, as one population replaces another due to repeated action of the mutant-selection event. The Theorem in its simplest form distinguishes between the two classes of resource constraints - limited, constant, and abundant, inconstant resources and admits the following description.

\begin{enumerate}
\item[II(a)] Limited, diverse resources: Evolutionary entropy increases.
\item[II(b)] Abundant, singular resources: Evolutionary entropy decreases.
\end{enumerate}

Directionality in these instances does not implicate population size and refers to the change in entropy, as the system evolves from one steady state to the next.

\section{Evolutionary Dynamics and Directionality Theory}
The analytic basis for the directionality principle described in \cite{D2} will be reviewed in this section. We refer to \cite{D1} and \cite{D2} for the detailed description of the concepts and mathematical arguments that underlie Eq. \eqref{eq9}.

\medskip
The Darwinian theory of evolution provides a necessary and sufficient mechanism for the adaptation of a population to its environment. The critical elements of the theory are inherent in three principles, see Levins and Lewontin  \cite{LL} and Demetrius  and Gundlach \cite{DG1}.
\begin{enumerate}
	\item[(i)] Physiological, behavioral and morphological traits vary among the members of a population (\textit{Variation})
	\item[(ii)]	The phenotypic traits are partly heritable: Descendants in a lineage will have traits similar to their ancestors (\textit{Heritability}).
	\item[(iii)] Different variants have different capacities to appropriate resources from the external environment	and to convert these resources into metabolic energy, and the demographic currency of survivorship and reproduction (\textit{Natural Selection})
\end{enumerate}	
These principles entail that the frequency of different types in a population will change due to the continued generation of new variants, and  selection against those types who are less effective in dealing with the exigences of the environment.

\ms
The dynamical system  which characterizes these three principles can be described as a two-step process: 
\begin{enumerate}
	\item [(a)]\textbf{ Mutation; } Random changes  in the genetic endowment of a small subset of the incumbent population.\\
	These changes will result in the incidence of two types  --- the incumbent, endowed with the ancestral genotype, and the variant with a  mutant inheritance.
	\item [(b)] \textbf{Selection:} Competition between the incumbent and the variant types for the resources  of the external environment.
\end{enumerate} 
Directionality Theory is the study of the dynamical system generated by the process of mutation and natural selection. The analysis assumes that the fundamental unit of the evolutionary process  is a population. This biological object is an aggregate of replicating  units. The evolutionary argument applies to units at different hierarchical levels:  macromolecules, cells,  organisms.

\ms
Formally, a population can be described by a directed graph, denoted by $G$, see Fig. (3)

The nodes of the graph represent the replicators. The links between the nodes describe the interaction between the different 
agents.

\ms
At steady state, the population process, depicted as a directed graph, can be represented as an abstract dynamical system, defined in terms of the following elements:
\begin{enumerate}
	\item[(i)]  $\Omega$: The set of genealogies, that is the set of sequences generated by the interaction between the elements
	\item[(ii)]  The probability measure $\mu$. This  parameter describes the frequency distribution of the genealogies.  
	\item[(iii)] The shift operator $\sigma$ defined by
	\[\sigma:(x_k)_k\mapsto (x_{k+1})_k. \]
	
	\end{enumerate} 
	The element $x=(\cdots x_{-2}x_{-1}x_0x_1x_2\cdots)$ is a genealogy.
	
	The probability measure $\mu$ described in (ii) is the steady state, the Gibbs measure induced by the potential function $\phi:\Omega\to \bR$.

\bigskip
\subsection{Dynamical systems and Evolutionary Entropy}

\smallskip
We consider the population as an abstract dynamical system of the form $(\Omega,\mu,\sigma)$, where $\sigma$ denote the shift map on $\Omega$. Two measure preserving transformations $(\Omega,\mu,\sigma)$, $(\Omega^*,\mu^*,\sigma^*)$ are said to be isomorphic if there is a one--to--one correspondence between all (but a set of measure zero) of the points in each measure space, so that corresponding points are transformed in the same way.

\ms
The dynamical entropy, the Kolmogorov--Sinai entropy, constitutes an isomorphism invariant of measure preserving transformations. This mathematical object is thus a fundamental statistical invariant of the dynamical system.

\ms
The work initiated in Demetrius \cite{D3} exploited the notion of isomorphism invariant of measure preserving transformations to develop an intrinsic and fundamental property of the population dynamics. This quantity is called \textit{evolutionary entropy.}
The term "entropy" refers to the mathematical origin of the concept: in particular its significance in the classification of measure preserving transformations. 
The term "evolutionary" reflects the biological roots of the concept. The statistical measure predicts the outcome of competition between an incumbent population and a variant type,  and hence constitutes a quantitative measure of Darwinian fitness \cite{D3}.

\ms

The concept was originally introduced in the study of the evolutionary dynamics of demographic networks. These systems can also be described in terms of a directed graph. The nodes of the graph correspond to age classes, the interaction between the nodes represents the flow of energy from one age class to another. This energy flow is in terms of survivorship from one age-class to the next; and reproduction, an energy flow from the  age-class of reproductives to the age class of newborns. These processes can be represented in terms of an abstract dynamical system $(\Omega,\mu,\phi)$. The function $\phi:\Omega\to \bR$ which describes the survivorship and reproduction schedule is locally constant \cite{D3}. The evolutionary entropy $H$, in these classes of models can be expressed in the form
\begin{equation}
H=\frac{\widetilde{S}}{\widetilde{T}}.
\end{equation}
Here $\widetilde{S}$ is a measure of the uncertainty in the age of the model of a randomly chosen newborn and $\widetilde{T}$ is generation time.

\ms

There exist many biological systems in which the individual elements can be parameterized in terms of a finite number of classes. These include the demographic models we will discuss in this article. In these systems, the phase space $\Omega$ can be modeled by a subshift of finite type with a general transition matrix $A$ and a potential function $\phi:\Omega\to\bR$ that is locally constant. This locally constant condition excludes many phenomena of scientific interest, for example bioenergetic processes which  
exist in metabolic reactions, and the interaction which describe the exchange of non-material resources in social networks.
\subsubsection{Continuous Potentials}

We now develop a general theory for the evolutionary entropy for \textit{continuous } potentials, to accommodate these processes. We restrict our representation to the introduction of  the relevant objects and results, see e.g. \cite{Aa},\cite{B},\cite{Ki},\cite{MU},\cite{R},\cite{W}.

\ms
We will describe as before, the population as a mathematical object -- a directed graph as depicted in Fig. (4). The nodes of the graph represent the states, namely groups of individuals of a given age or size, as in the analysis of demographic networks \cite{D3}, or groups defined in terms of their behavior or social norms, as in the analysis of social networks .  The links between the nodes describe the transfer of energy between the various states.

\ms
Let
$$
X=\{1,\dots,d\}
$$
and
$$
Y=\prod\limits^\infty_{n=-\infty} X_n
$$
where $X_n=X$. Let $A=(a_{ij})_{1\leq i,j\leq d}$ be a $d\times d$-matrix with entrees in $\{0,1\}$.
Let 
\begin{equation}\label{defSFT}
\Omega=\Omega_A=\{x\in Y : a_{x_ix_{i+1}}=1\}
\end{equation}
and let  $\sigma:\Omega \to \Omega$ denote the shift operator on $\Omega$. We say $(\Omega,\sigma)$ is a subshift of finite type with transition matrix $A$.  Let $\phi:\Omega\to \bR$ be a continuous potential. 
\ms

Our goal is to establish a formula for the \emph{Evolutionary entropy} $H_{{\rm evol}}(\phi)$ (which we will simply denote by $H$) in terms of a quantity that determines the diversity in pathways of energy flow denoted by $\widetilde{S}$, and the mean cycle time $\widetilde{T}$.
\ms

 Let $\cM$ denote the set of all $\sigma$-invariant Borel probability measures on $\Omega$, and let $\cM_E\subset \cM$ denote the subset of ergodic measures, see e.g. \cite{W} for the definitions.
\ms

Given $n\geq 0$ we say an $n$-tuple $\tau=\tau_0\cdots \tau_{n-1}\in X^{n}$ is $\Omega$-admissible provided that $A_{ab}=1$ for all pairs of consecutive elements $ab$ in $\tau$. We denote by $\cL_\Omega^n$ the set of all $A$-admissible tuples of length $n$.
Given $\tau\in \cL_\Omega^n$ we denote by 
\[
[\tau]=\{x\in \Omega: x_0=\tau_0,\dots, x_{n-1}=\tau_{n-1}\}\]  
the cylinder of length $n$ generated by $\tau$.

\bigskip
\subsubsection{The growth rate parameter and the variational principle}
\ms
 Next we introduce the growth rate parameter of the  potential $\phi$.  Given $n\geq 1$ we define the $n$-th \emph{partition function} $Z_n(\phi)$ at $\phi$ by
\begin{equation}\label{eqzn}
Z_n(\phi)=\sum_{\tau\in \cL^n_\Omega} \exp\left(\sup_{x\in [\tau]} S_n\phi (x)\right),
\end{equation}
where
\begin{equation}\label{eqsn}
S_n\phi (x)=\sum_{k=0}^{n-1} \phi(\sigma^k(x))
\end{equation}
denotes  statistical sum of length $n$ of $x$.  We define the \emph{growth parameter}\footnote{Note that in the context of dynamical systems $r(\phi)$ is refered to as the topological pressure  $\Ptop(\phi)$ of the potential $\phi$ \cite{MU,R,W}.} of $\phi$ (with respect to the shift map $\sigma$) by
\begin{equation}\label{defptop}
r(\phi)=\lim_{n\to \infty} \frac{1}{n} \log Z_n(\phi)=\inf \left\{\frac{1}{n} \log Z_n(\phi): n\geq 1\right\}.
\end{equation}
Moreover, $\htop(\sigma)=r(0)$ denotes the {\em topological entropy} of $\sigma$. Recall that  $\htop(\sigma)=\log \lambda$ where $\lambda$ is the spectral radius of the transition matrix $A$. 
Given $\mu\in \cM$ we denote by $h_\mu(\sigma)$ the Kolmogorov-Sinai entropy of the measure $\mu$ given by 
 \begin{equation}\label{defmeasent}
 h_\mu(\sigma)=\lim_{n\to\infty}-\frac{1}{n}\sum_{\tau\in \cL^n_\Omega} \mu([\tau])\log \mu([\tau]),
 \end{equation}
 where terms with  $\mu([\tau])=0$ are omitted from the sum.
  The growth rate parameter satisfies the well-known \emph{variational principle} (see e.g. \cite{W}):
 \begin{equation}\label{varpri}
 r(\phi)=\sup_{\mu\in \cM} \left(h_\mu(\sigma)+\int\phi\, d\mu\right).
 \end{equation}
 If $\mu\in\cM$ achieves the supremum in \eqref{varpri},   we call $\mu$  an \emph{equilibrium state} of $\phi$. We denote the set of equilibrium states of $\phi$ by $ES(\phi)$. Recall that $ES(\phi)$ is nonempty. 

\bigskip

\subsubsection{Evolutionary Entropy and Cycle times}
\ms
 We define the  \textit{evolutionary entropy} of $\phi$ by
\begin{equation}\label{defEE}
H=H(\phi)=\sup\left\{h_\mu(\sigma) : \mu\in ES(\phi)\right\}.	
\end{equation}
We observe that the set  
\begin{equation}
\cR_{\phi}= \left\{\smallint \phi\, d\mu:\mu \in ES(\phi)\right\}
\end{equation}
 is a closed interval $[a_\phi,b_\phi]$. For $\mu\in ES(\phi)$ it follows from  \eqref{varpri} that 
 \begin{equation}
 H=h_\mu(\sigma)
  \end{equation}
  if and only if $\int\phi\, d\mu=a_\phi$. In particular, the supremum in equation \eqref{defEE} is a maximum. Further, by using ergodic decompositions combined with a convexity argument we conclude that there exists at least one $\mu\in ES(\phi)\cap \cM_E$ with 
 $H=h_\mu(\sigma)$. It turns out that for a "large" set of potentials $\phi$ the set $\cR_\phi$ is a singleton. For example, if $\phi$ is H\"older continuous then $ES(\phi)$ and  $\cR_\phi$ are singletons, see \cite{Bo1}.
 
 \ms
 
Let $(\Omega,\sigma)$ be a transitive subshift of finite type, and let $\phi:\Omega\to\bR$ be a continuous potential. Then the evolutionary entropy $H$  of $\phi$ is given by the formula
\begin{equation}\label{eqmain}
H=\frac{\widetilde{S}}{\widetilde{T}},
\end{equation}
where $\widetilde{S}$ is the limit of entropies  $\widetilde{S_n}$ of countable Bernoulli shifts  (see e.g. \cite{SO}) and $T$ denotes the mean cycle time of the system. In particular, if  $\phi$ is locally constant, then
$\widetilde{S}$ is the entropy of a countable (finite or infinite) Bernoulli shift. 
\ms 

We briefly discuss  the main differences between the theory of locally constant and continuous potentials. The locally constant case has been successfully applied to situations where the population can be partitioned into homogenous groups whose statistical properties are encoded by a small number of parameters. A prototype of such an example is the demographic model which we will discuss in the Section 4.2. In this model, the partition is obtained by dividing the population into  age-groups. We then  associate with each of these groups the probability of survivorship,  and with the reproductive classes, the mean number of offsprings produced by individuals in the group.

However, in  more heterogeneous situations the application of locally constant potentials may lead  to less accurate and in some cases misleading predictions. This is for example the case in the evolution of social behavior of humans. Social preferences and dispositions are continuous variables and individuals often invoke their memory of past interactions in deciding whether to cooperate or defect in social encounters \cite{BG}.
These situations can not be accurately modeled by partitioning.   

\ms

While the two mathematical approaches, the use of locally constant and continuous potentials, both originate in the mathematical thermodynamic formalism \cite{R}, they are based on different methods. Namely, in the locally constant case, the input is a finite set of parameters which yield an explicit formula for the Evolutionary entropy. In contrast, for continuous potentials, the Evolutionary entropy is implicitly defined and thus there is no explicit formula. It should be noted however  that it has been recently established that for a large class of continuous potentials the Evolutionary entropy is computable in the sense of computable analysis \cite{BW}, i.e., it can be computed by a Turing machine (a computer program for our purposes) at any pre-described accuracy.

\medskip
\subsubsection{ Demographic networks and evolutionary entropy}

\smallskip
Demographic networks whose behavior can be analyzed in terms of locally constant potentials constitute a well studied example where the evolutionary entropy concept can be explicitly characterized.

\ms
The network can be described by the directed graph given by Fig. (7)

\smallskip
\begin{center}

\begin{center}
\begin{tikzpicture}
\node at (0,0) (A) {\footnotesize{1}};
\node at (2,0) (B) {\footnotesize{2}};
\node at (4,0) (C) {\footnotesize{3}};
\node at (5.2,0) (D) {};
\node at (6.8,0) (De) {};
\node at (8,0) (E) {\footnotesize{d}};
\node at (-1.1,1.95)(F){$m_1$};
\draw[thick] (A) circle (.2);
\draw[thick] (B) circle (.2);
\draw[thick] (C) circle (.2);
\draw[thick] (E) circle (.2);

\begin{scope}[rotate=45]
\draw[thick](A) edge  [->,in=120,out=60,loop,{looseness=25}]  (A);
\end{scope}

\draw[thick] (A) edge[->] node[below]{$b_1$} (B);
\draw[thick] (B) edge[->] node[below]{$b_2$} (C);
\draw[thick] (C) edge[->] node[below]{$b_3$} (D);
\draw[thick] (De) edge[->] node[below]{$b_d$} (E);
\draw[thick] (B) edge[->,in=30,out=150] node[above right]{$m_2$} (A);
\draw[thick] (C) edge[->,in=55,out=125] node[above right]{$m_3$} (A);
\draw[thick] (E) edge[->,in=75,out=105] node[above right]{$m_d$} (A);

\draw[ultra thick, loosely dotted] (D)  --   (De);

\end{tikzpicture}
\end{center}

	Fig. (7)
\end{center}

The graph represents a population whose members are classified in terms of age classes.

\ms
The parameters $(b_i)$ denote the survivorship from age class $(i)$ to $(i+1)$. The parameters $(m_i)$ denote the mean number offspring produced by individuals in age class $(i)$.

\ms
The interaction matrix is given by
$$
A=\left[\ba{llll}
m_1&m_2&\ldots \ldots&m_d\\
b_1&0&\ldots \ldots&0\\
.& & &\\
.& & &\\
.& & &\\
0&0& \ldots \, \, \, \, \, \,\, \, b_{d-1}&0
\ea\right]
$$

The potential $\phi : \Omega\to \bR$ will be locally constant and described by $\phi(x)=\log a_{x_0x_1}$.

\ms

Hence Evolutionary entropy, $\EE(\phi)$, can be obtained by evaluating the entropy $H$ of the Markov chain.

$$
P=\left[\ba{llll}
p_1&p_2&\ldots \ldots&p_d\\
1&0&\ldots \ldots&0\\
0&1&\ldots \ldots&0\\
.& & &\\
.& & &\\
.& & &\\
0&0&\ldots\, \, \, \, \, \, \, \,1&0
\ea\right]
$$
where 
$$
p_j=\frac{l_jm_j}{\lambda^j}
$$	
The parameter $\lambda$ is the dominant eigenvalue of the matrix $P$. The function $l_j$ is given by
$$
l_j=\left\{\ba {ll}
1&j=1\\
b_1\ldots b_{j-1}& j > 1\ea\right.
$$
The entropy $H$ of the Markov chain is
\begin{equation}\label{eq20}
H=\frac{\widetilde{S}}{\widetilde{T}},
\end{equation}
where 
$$
\widetilde{S}= -\sum^d_{j=1}p_j\log p_j
$$
and 
$$
\widetilde{T}= \sum^d_{j=1}j p_j
$$
The quantity $\widetilde{S}$ is the variability in the age at which individuals reproduce and die and the
 quantity $\widetilde{T}$  denotes the mean cycle time. We note that \eqref{eq20} is  the special case of  \eqref{eqmain} for locally constant potentials.

\bigskip
\subsection{The Entropic Selection Principle}

The mathematical model of the Darwinian process considers factors such as the resource abundance and the resource variation as elements of the evolutionary process. The model postulates mutation, selection and inheritance as 
the principles underlying the dynamics of evolution.

The theory distinguishes between the incumbent population, described by a dynamical system $(\Omega,\mu,\phi)$ and a variant of small size described by a system $(\Omega,\mu(\delta),\phi(\delta))$. Here $\phi(\delta)$ is a small perturbation of $\phi$ of the form
\begin{equation}
\phi(\delta)=\phi+\delta \psi,
\end{equation}
where $\int\phi\, d\mu=\int \psi\, d\mu$.\footnote{We note that this condition does not restrict the generality of the approach since it  can be achieved by re-normalizing $\psi$.}
\ms

The \underline{Entropic Selection Principle} is concerned with the dynamics of the competition between the Incumbent population $(\Omega,\mu,\phi)$,  and the Variant $(\Omega,\mu(\delta),\phi(\delta)$. The analysis of the Entropic Selection Principle requires a coupling of the macroscopic parameter of the two dynamical systems \cite{D2}.

\noindent
(I) The parameters that describe the incumbent  $(\Omega,\mu,\phi)$  are (see Section 4.2):
\begin{enumerate}
\item[(1)] The Evolutionary entropy $H=\frac{\widetilde{S}}{\widetilde{T}}.$
\item[(2)] The growth rate $r(\phi)=\lim_{n\to\infty} \frac{1}{n}\log Z_n(\phi)$.
\item[(3)] The Demographic Index $\Phi=\lim_{n\to\infty} \frac{1}{n} E_n(S_n \phi)$.
\item[(4)] The Demographic Variance $\sigma^2=\lim_{n\to\infty} \frac{1}{n} Var_n(S_n \phi).$
\item[(5)] The Correlation index $\kappa=\lim_{n\to\infty} E_n[S_n\phi-E_n[S_n\phi]]^3$.
\end{enumerate}

The Evolutionary entropy $H$, the growth rate $r$, and the reproductive potential are related by the identity
\begin{equation}
r=H+\Phi.
\end{equation}

\noindent
(II) The Variant population $(\Omega,\mu(\delta),\phi(\delta)$ is derived from a mutation. This is represented in terms of a perturbation of the function $\phi$. We denote the parameters that characterize the variant population by 
$(r^\ast,H^\ast,\Phi^\ast,{\sigma^\ast}^2)$ where $r^\ast=r(\delta), H^\ast=H(\delta),\Phi^\ast=\Phi(\delta)$ and ${\sigma^\ast}^2=\sigma^2(\delta)$. Further, we define
\begin{equation}
\Delta r=r^\ast -r,\,\, \Delta H= H^\ast - H, \Delta\, \sigma^2={\sigma^\ast}^2-\sigma^2.
\end{equation}
We have, by using  perturbation methods in \cite{D2}, the following relations:
\begin{equation}\label{eq2888}
\Delta r =\Phi \delta,\,\, \Delta H = -\sigma^2\delta, \,\, \Delta \sigma^2=\gamma \delta,
\end{equation}
where $\gamma=2\sigma^2+\kappa$.

\subsubsection{Invasion--Extinction Dynamics}

\smallskip
The condition for the increase in frequency and ultimate fixation of the variant population is evaluated by considering the stochastic dynamics of the frequency of the variant.

The continuous time diffusion approximation is used in order to consider the function
\begin{equation}
p(t) = \frac{N^*(t)}{N(t)+N^*(t)}
\end{equation}
Here $N(t)$ denote the population size of the resident population and $N^*(t)$ the population size of the invader.

\ms
Let $f(N,t)$ and $f^*(N^*,t)$ denote the density of the processes $N(t)$ and $N^*(t)$. The evolution of the density is given by the solution of the Fokker--Planck equation
\begin{equation}
\frac{\partial f}{\partial t}=- r\frac{\partial(fN)}{\partial N}+\sigma^2\frac{\partial^2(fN)}{\partial N^2}
\end{equation}
and
\begin{equation}
\frac{\partial f^*}{\partial t} = - r^*\frac{\partial(f^*N^*)}{\partial N^*} + \sigma^{*2}\frac{\partial^2(f^*N^*)}{\partial N^{*2}}
\end{equation}
We can now invoke the constraint, total population size,
\begin{equation}
M=N(t)+N^*(t)
\end{equation}
$M$ constant, to derive a Fokker --Planck equation for the probability density function $\psi(p,t)$ of the stochastic process which describes the change in frequency of the invading population. We have
\begin{equation}\label{eq111}
\frac{\partial\psi}{\partial t}=-\frac{\partial[\alpha(p)\psi]}{\partial p}+\frac 12\frac{\partial^2(\beta(p)\psi]}{\partial p^2},
\end{equation}
where
\begin{equation}
\alpha(p)=-p(1-p)[\Delta r-\frac 1M\Delta\sigma^2]
\end{equation}
and
\begin{equation}
\beta(p)=\frac{p(1-p)}{M}[\sigma^2 p + \sigma^{*2}(1-p)]
\end{equation}
The analysis of \eqref{eq111} shows that the outcome of competition between the invading type and the resident population is determined by the selective advantage $s$, given by
\begin{equation}\label{eq3666}
s=\Delta r-\frac 1M\Delta\sigma^2
\end{equation}

The perturbation relation given by  \eqref{eq2888} entails the following implications.

\ms
$\Phi < 0 \Rightarrow \Delta r \cdot\Delta H > 0$\, ,\, \, \, $\Phi > 0 \Rightarrow \Delta r \cdot\Delta H < 0$\\
$\gamma < 0 \Rightarrow \Delta\sigma^2\Delta H > 0$, \, \, 
$\gamma> 0 \Rightarrow \Delta\sigma^2 \cdot\Delta H < 0$

\ms
We can now apply these relations to express the selection advantage \eqref{eq3666} in the form
\begin{equation}
s^*=-(\Phi-\gamma/M)\Delta H
\end{equation}
The relation between $\Phi,\gamma$, and the change $\Delta H$ is given by
\begin{equation}\label{eq30}
(-\Phi+\gamma/M)\Delta H > 0
\end{equation}

The relation \eqref{eq30} is the \underline{Entropic Selection Principle}: This principle asserts that the outcome of competition between
the incumbent and the variant population is contingent on the parameters $\Phi$ and $\gamma$ and determined by $H$.

are summarized in Table (1) and (2).

\begin{center}
		
	\smallskip
	\begin{tabular}{|l|l|}
		\hline
		Constraints $\Phi,\gamma$&Outcome\\
		\hline
		$\Phi < 0$,\, \, $\gamma > 0$&$\Delta H > 0$\\
		$\Phi > 0$,\, \, $\gamma < 0$&$\Delta H < 0$\\
		\hline
	\end{tabular}
	\vspace{0.6cm}
	\ms
	
	\textbf{Table 1} Relation between macroscopic parameters $\Phi, \gamma$ and Selection Outcome, $\Delta H$

	\bigskip
	
	\smallskip
	\begin{tabular}{|l|l|l|}
		\hline
		Constraint&Population Size&Outcome\\
		\hline
		& $M> \frac{\Phi}{\gamma}$&$\Delta H > 0$\\
		$\Phi < 0,\, \gamma < 0$& & \\
		& $M<\frac{\Phi}{\gamma}$ &$\Delta H < 0$\\
		& & \\
		& $M> \frac{\Phi}{\gamma}$&$\Delta H < 0$\\
		$\Phi > 0,\, \gamma > 0$& & \\
		&$M< \frac{\Phi}{\gamma}$&$\Delta H > 0$\\
		\hline
		
		\end{tabular}
		\vspace{0.6cm}
		\ms 
		
		\textbf{Table 2} Relation between macroscopic parameters $\Phi ,\gamma$ and Selection Outcome $\Delta H$
\end{center}

\section{Statistical Thermodynamics and Evolutionary Theory}

Statistical Thermodynamics is concerned with understanding the macroscopic properties of matter in terms of the molecular constituencies. The fundamental concepts in this discipline
are temperature (i.e. the mean kinetic energy of the molecules), and the thermodynamic entropy, i.e., the number of molecular configurations which are associated with a given macrostate.

Evolutionary dynamics is concerned with  understanding  the flow and transformation of energy in populations of replicating organisms in terms of the dynamical behavior and the birth and death rates of the 
individuel organisms.

The fundamental parameters in this theory are the mean cycle time, i.e. the generation time, and the Evolutionary entropy, i.e., the number of replicating cycles generated by the interaction between the individuals.

We show that these two theories are isomorphic in the sense that there exists a correspondence between the macroscopic parameters that define the theories.

We will furthermore show in Section 6 that the correspondence between the classes of variables is analytic. We will explore this analyticity to show that the Fundamental Theorem of  Evolution, the directionality principle for evolutionary entropy, 
is a generalization of the Second Law of Thermodynamics, the directionality principle for thermodynamic entropy.

\subsection{Statistical mechanics of a gas}
We consider a gas as a system consisting of N interacting molecules. Let $X=\{1,\dots,d\}$ denote the phase space. Further let $\cM$ denote the set of probability measures on $X$. Consider a potential function $\phi:X\to \bR$ as representing the potential energy. The mean energy of the system in state $\mu=(\mu_i)$ is given by
\begin{equation}
\Phi=\sum_{i=1}^d \mu_i \log \phi(x_i)=\mu(\log \phi).
\end{equation}
The Entropy $S(\mu)$ is given by
\begin{equation}
S(\mu)=-\sum \mu_i \log \mu_i.
\end{equation}
The quantity $Z$ is determined by the variational principle 
\begin{equation}\label{eqgasvarpri}
\log Z =\sup_{\mu\in \cM} \left[ \mu(\log \phi)+ S(\mu)\right].
\end{equation}
Moreover, the maximum in \eqref{eqgasvarpri} is attained by a unique measure $\hat{\mu}$, that is, $\log Z =\hat{\mu}(\log \phi)+ S(\hat{\mu})$.
The physical interpretation ascribed to the variational principle can be discerned by expressing the potential function in the form
\begin{equation}
\phi(x_i)=\exp(-\beta E_i),
\end{equation}
where $\beta=\frac{1}{k T}$. The quantity $T$ is the temperature and $k$ is the Boltzmann constant. The expression for the distribution $\hat{\mu}=(\hat{\mu}_i)$ now becomes
\begin{equation}
\hat{\mu}_i=\frac{\exp(-\beta E_i)}{\sum \exp(-\beta E_i)}.
\end{equation}
The variational principle then asserts that the distribution $\hat{\mu}$ maximizes $S-\frac{E}{k T}$. Equivalently it minimizes free Energy $ 
F$ which is given by
\begin{equation}\label{eq4444}
F=E-k\, S\,T.
\end{equation}

\subsection{Statistical mechanics of a population}
We consider an age-structured population whose dynamics is given by
\begin{equation}\label{eq4555}
\tilde{u}(t+1)= A\, \tilde{u}(t),
\end{equation}
where $ \tilde{u}(t)$ denotes the age-distribution and $A$ the population matrix
\begin{equation}
A=\left[\ba{llll}
m_1&m_2&\ldots \ldots&m_d\\
b_1&0&\ldots \ldots&0\\
.& & &\\
.& & &\\
.& & &\\
0&0& \ldots \, \, \, \, \, \,\, \, b_{d-1}&0
\ea\right]
\end{equation}
Let $\Omega$ denote the phase space, i.e., the set genealogies generated by the graph associated with the matrix $A$. Let $\widetilde{\cM}$ denote the space of shift-invariant probability measures on $\Omega$.

Consider the function 
\begin{equation}
\phi(x)=\log a_{x_0,x_1}.
\end{equation}
The mean energy is given by
\begin{equation}
\Phi(\mu)=\int \phi\, d\mu.
\end{equation}
The entropy $H_\mu(\sigma)$ is the Kolmogorov-Sinai entropy of the measure $\mu$, see \eqref{defmeasent} for the definition. With these definitions the following variational principle holds:
\begin{equation}
\log \lambda= \sup_{\mu\in \widetilde{\cM}}\left[ H_\mu(\sigma)+\int \phi\,d\mu\right].
\end{equation}
The supremum is attained by a unique $\tilde{\mu}\in  \widetilde{\cM}$. Hence
\begin{equation}\label{eq505050}
\log \lambda=H+\Phi\quad {\rm and }\quad \log \lambda=\Phi+\frac{\widetilde{S}}{\widetilde{T}},
\end{equation}
where 
\begin{equation}
\widetilde{S}=-\sum p_j\log p_j,\,\, \Phi=\frac{\sum p_j \log \phi_j}{\widetilde{T}}\, \, {\rm and}\,\, \widetilde{T}=\sum j\phi_j.
\end{equation}
\subsection{Relations between the Macroscopic Parameters}
We can use the expressions given in \eqref{eq4444} and \eqref{eq505050} to derive a formal relation between the two classes of macroscopic parameters. This correspondence is given in Table 1.

\begin{center}
		
	\smallskip
	\begin{tabular}{|l|l|}
	\hline
	
		Thermodynamic variable & Evolutionary Parameters\\
		\hline
		Free Energy $F$ & Growth rate $\tilde{r}$\\
		Inverse Temperature $T$ & Generation Time $\tilde{T}$\\
		Mean Energy $E$ & Reproductive Potential $\Phi$ \\
		Thermodynamic Entropy $S$ & Population Entropy $ \widetilde{S}$\\
		\hline

			\end{tabular}
	\vspace{0.6cm}
	\ms
	
	\textbf{Table 3} Relation between the macroscopic parameters
\end{center}

\section{Directionality Principles: Thermodynamics and Evolutionary Theory}

The dynamics of molecular collisions in a macroscopic body are encoded by thermodynamic entropy.
The dynamics of the lower level interactions in a population of replicating organisms are encoded by Evolutionary entropy. These two parameters, as shown in Section 5, are formal analogues. We will now show that they 
are analytically related. This relation will be the cornerstone for the analytical fact that the Fundamental Law of Evolution is a generalization of the Second Law of Thermodynamics.

\subsection{Energy transformation: Inanimate matter}

\smallskip
Energy transformation in inanimate matter is determined by the Second Law of Thermodynamics. The Law asserts that Thermodynamic entropy $S$ increases. 

\ms
We write
\begin{equation}\label{eq2law}
\Delta S > 0,
\end{equation}
where
\begin{equation}\label{eqW}
S=k_b\log W.
\end{equation}
In equation \eqref{eqW} the quantity $S$ describes the extent to which energy is spread and shared among the microscopic energy modes of the system whereas
the parameter $W$  denotes the number of molecular configurations that are compatible with the macrostate of the system. 
 The validity of inequality \eqref{eq2law}  requires that the system is isolated and closed to the input of energy and matter.

\bigskip
\subsubsection{Energy transformation in living matter}

\smallskip
The evolution of energy in living matter is determined by the Entropic Selection Principle.

\ms
This principle applies to systems which are open to the input of energy and matter.

\ms
Recall that by  \eqref{eq30} the Evolutionary entropy $H$ evolves according to the rule
\begin{equation}\label{eq301}
(-\Phi+\gamma/M) \Delta H > 0.
\end{equation}
Now let $R$ denote the Resource endowment and assume that $R$ evolves according to the differential equation
\begin{equation}
d\, R(t) = \rho\,  R(t)dt +\beta\, dt,
\end{equation}
where $\rho$ denotes the production rate of the external resource.
We now assume that the Resource process and the population process are in dynamical equilibrium. Here we use
\begin{equation}
\frac{adr(\delta)}{d\delta}\Bigg|_{\delta = 0} = \Phi, \quad \frac{ad\sigma^2(\delta)}{d\delta}\Bigg|_{\delta = 0} = \gamma
\end{equation}
and write $\rho = a\Phi$, $\beta=a\gamma$.

\ms
If we assume that there is no exchange in energy and matter between the population and the external environment, we have
\begin{equation}
\rho=0,\quad \beta=c,
\end{equation}
which implies that
\begin{equation}
\Phi=0\quad\mbox{\rm and}\quad \gamma=k
\end{equation}
where $k$ is a numerical constant.

\ms
The Laws describing the changes in evolutionary entropy will in view of \eqref{eq301} become
\begin{equation}
\Delta H > 0,
\end{equation}where 
\begin{equation}
H=\frac{\widetilde{S}}{\widetilde{T}}.
\end{equation}
Since
\begin{equation}
\Delta H\cdot\Delta \widetilde{S} > 0
\end{equation}
we conclude that the directionality principle, with constraints on the resource endowment, is given by
\begin{equation}
\Delta \widetilde{S} > 0.
\end{equation}

\bigskip
\subsubsection{Energy transformation: inanimate matter, living matter}

\smallskip
Irreversibility in the Second Law of Thermodynamics is given in terms of the function
$$
\Delta S > 0
$$
where $S$ denotes the thermodynamic entropy.

\ms
Irreversibility in the Fundamental Principle of Evolution is given by 
$$
(-\Phi+\gamma/M)\Delta H > 0.
$$
We have shown that when the system is transformed from a process open to input of energy and matter to an isolated system,  \eqref{eq30} reduces to the condition
$$
\Delta \widetilde{S} > 0
$$
We will now establish a relation between thermodynamic entropy $S$ and evolutionary entropy $H$. 

\ms
We first recall that
\begin{equation}\label{eq45}
H=\frac{\widetilde{S}}{\widetilde{T}},
\end{equation}
where 
\begin{equation}
\widetilde{S}=-\sum_k p_k\log p_k,
\end{equation}
and
\begin{equation}
\widetilde{T}=\sum_k k \, p_k.
\end{equation}
Let $N$ denote the population size, which is assumed to be large. We also assume that the total number of replicative cycles is also of the order $N$. Hence $n_k$, the number of cycles of length $k$, is of the order $n_k=p_k N$

We write 
\begin{equation}
p_k=\frac{n_k}{N},
\end{equation}
where 
\begin{equation}
N=\sum n_k.
\end{equation}
Therefore, $p_k$ is the probability that a randomly chosen cycle in the network of interactive microstates has length $k$.

Therefore,
\begin{equation}
\widetilde{S}= -\sum_k\frac{n_k}{N }\log \frac{n_k}{N }.
\end{equation}
Hence
\begin{equation}
\widetilde{S}= -\frac{1}{N} \sum n_k\log\left(\frac{n_k}{N}\right),
\end{equation}
and 
\begin{equation}
\widetilde{S}=-\frac {1}{N} \sum_k n_k(\log n_k-\log N).
\end{equation}
We conclude that
\begin{equation}
N\, \widetilde{S}=-\sum_k n_k\log n_k + \sum_k n_k\log N, 
\end{equation}
which implies
\begin{equation}\label{eq59}
N\, \widetilde{S}= N \log N -\sum_k n_k\log n_k.
\end{equation}
Recall that by Stirling's formula we have
\begin{equation}\label{eq60}
\log N! = N\log N - N.
\end{equation}
Combining \eqref{eq59} and \eqref{eq60} yields
\begin{equation}
N \, \widetilde{S} = \log\left[\frac{N !}{n ! n_2\ldots}\right]
\end{equation}
We conclude that
\begin{equation}
\widetilde{S}=\frac 1 N \log \left[\frac{ N !}{
n_1! n_2!\ldots}\right]
\end{equation}
Now let $W$ denote the number of microscopic states which are compatible with a given macrostate. 
The thermodynamic entropy $S$ is given by
\begin{equation}
S= k_b\log W,
\end{equation}
where
\begin{equation}
W=\left[\frac{ N !}{n_1!  n_2!\ldots}\right].
\end{equation}
Hence 
\begin{equation}
S=k_B\tilde{S}
\end{equation}
We conclude that as $\rho\to 0$, the thermodynamic entropy and the evolutionary entropy coincide.

\ms
The relation between the two measures of organization implies that the Fundamental Theorem of evolution is a generalization of the Second Law of thermodynamics.

\ms
The relation between the parameters which are involved in energy transformation in physical and biological systems is described in Table (3).

\bigskip
\begin{center}
	\textbf{Table 4} Relation between the parameters in Thermodynamic\\Theory and Evolutionary Theory.
	
	\medskip
	\begin{tabular}{lll}
		\hline
		\textbf{Parameter}&\textbf{Thermodynamic Theory}&\textbf{Evolutionary Theory}\\
		\hline
		Organizing Variable&Temperature&Cycle Time\\
		Fitness Parameter&Thermodynamic Entropy $S$&Evolutionary Entropy $H$\\
		Selection Principle&$\Delta S > 0$&($-\Phi + \gamma/M)\Delta H > 0$\\
		\hline
	\end{tabular}
\end{center}

\section{Discussion}

The two classes of entities that constitute the natural world - the aggregates of inanimate matter, and the populations of living organisms, both manifest a hierarchical structure with ordering in terms of time and energy scales.
The aggregates in the physical world - the ensemble of solids, liquids and gases, range from the submicroscopic to galactic. The elements in the living world, the integrated assembly of DNA, RNA and proteins, scale from viruses, to uni-cells, to multi-cells and to communities of plants and animals. Complex human societies, organized by both genes and culture, are at the top of this hierarchy. 

The various states of organization in aggregates of inanimate matter, and in populations of cells and higher organisms are the outcome of the transfer and transformation of energy. Energy is a collective concept which can exist in many 
forms: The most commonly encountered forms are heat (thermal energy), motion (kinetic or mechanical energy), light (electromagnetic energy) and metabolism (chemical energy).

The laws which pertain to energy transformation in inanimate matter are based on the empirical observation that in isolated systems, that is, systems closed to the input of energy and matter, there is the tendency for energy to disperse and spread within
the enclosure \cite{At}, \cite{Be}.

The molecular dynamics explanation of this principle can be formulated in terms of the analytical rule
\[
\Delta S\geq 0.
\]
 The quantity $S$, a measure of positional disorder, is given by
 \[ 
 S=k_B \log W,
 \]
 where $W$ denotes the number of microscopic configurations consistent with a given macrostate and $k_B$ is the Boltzmann constant. 
 
 Energy transformations in populations of metabolic and replicating entities: macromolecules, cells, higher organisms, occur under constraints which are quite distinct from the typical situations observed in aggregates of inanimate matter. There are three characteristic aspects:
 
 \begin{itemize}
 \item[(i)] \underline{Openness}: Living organisms are maintained by a continuous exchange of energy and matter with the external environment.
 
 \item[(ii)] \underline{Isothermal condition}: The low temperature differential between the organelles in a cell indicate that the cells do not act as heat engines.  Living organism are isothermal chemical machines.
 
 \item[(iii)] \underline{Size}: The number of molecules in a cell, and the number of cells in a population are of magnitude much smaller than the number of molecules in a gas.
 \end{itemize}
 
 These constraints entail that evolutionary selection, the process that drives the transfer and transformation of energy in populations of replicating organisms, will necessarily have a different character from thermodynamic selection, the 
 process describing energy transformation in aggregates of inanimate matter.
 
 The mathematical analysis of  thermodynamic processes shows that thermodynamic entropy, a measure of positional disorder, will be replaced by evolutionary entropy, a measure of temporal
 organization.
 
 Evolutionary entropy describes the rate at which the population appropriates  chemical energy from the external environment and converts this energy into biological work.
 Evolutionary entropy, H is given by
 \begin{equation}
 H=\frac{\widetilde{S}}{\widetilde{T}}.
 \end{equation}
 The quantity $\widetilde{S}$ denotes the number of bioenergetic cycles in a population of metabolic and replicating enteties. The quantity $\widetilde{T}$ denotes the mean cycle time.
 
 In the evolutionary process, directional changes in evolutionary entropy, will be contingent on the external resource constraint. These changes are expressed by
 \begin{equation}\label{fin}
 \left(-\Phi+\frac{\gamma}{M}\right)\Delta H >0.
 \end{equation}
The quantities $\Phi$ and $\gamma$ are correlated with the resource endowment, its amplitude and its variability, respectively.

The relation \eqref{fin}, the kernel of the Fundamental Theorem of Evolution, entails that:
 \begin{itemize}
 \item[(i)] Evolutionary entropy increases when the resource endowment is scarce and diverse.
 \item[(ii)] Evolutionary entropy decreases when the resource endowment is abundant and singular.
 \end{itemize}

 The directionality Principle, as given in \eqref{fin}, is applicable to the energy transformation at various 
 hierarchical levels.
 
 (1) \underline{Molecular:} The principle has provided an explanation for the changes in sequence length observed in experimental studies of the evolution of the Q$\beta$ virus. These studies show that sequence length increases when the resource is scarce, and decreases when the resource abundant, see, e.g. \cite{Sp}.
 
 (2) \underline{Demographic:} The evolution of life history: The principle elucidates the  increase in iteroparity when the resources are scarce, and the shift to semelparity when they are abundant, see \cite{St}, \cite{ZD}.
 
 (3) \underline{Social:} The evolution of cooperation: Cooperation in a social network refers to the interaction between social agents to achieve a particular goal. This activity may involve costs and benefits. The principle predicts altruistic
  behavior   when the resources are scarce and diverse, and selfish behavior when the resources are abundant and singular \cite{DG1}.
 \ms

 This article has shown that the Fundamental Theorem of Evolution is the natural generalization of the Second Law of Thermodynamics.
 
 Both Laws are concerned with Energy and its Transformation. The Laws, however have different domains of validity. This fact derives from the different constraints that regulate energy transformation in inorganic matter and living organisms.


\begin{thebibliography}{99}

\bibitem{Aa}Aaronson, J: \emph{An introduction to infinite ergodic theory,} Mathematical Surveys and Monographs {\bf 50}, AMS, 1997.
	\bibitem{At} Atkins, P.: \emph{Conjuring the universe: The origins of the laws of nature}, Oxford University Press, 2018.
	\bibitem{Be}  Berry F.S.:    \emph{Three laws of nature}, Yale Univ. Press, 2019.
	\bibitem{B} Bowen, R.:\textit{ Equilibrium states and the ergodic theory of Anosov diffeomorphisms}, Second revised edition. With a preface by David Ruelle. Edited by Jean--Rene Chazottes. Lecture Notes in Mathematics, 470. Springer Verlag, Berlin 2008, viii+75 pp.
	\bibitem{Bo1} Bowen, R.: \textit{Some systems with unique equilibrium states}, Math. Systems Theory {\bf 8}, (1974/1975), 193--202.	
	\bibitem{BG}Bowles, S and Ginitis, H., \emph{A cooperative species: Human reciprocity and its evolution}, Princeton University Press, 2011. 
	\bibitem{BW} 	Burr, M. and Wolf, C., \textit{Computability at zero temperature}, preprint.
	\bibitem{D1} Demetrius, L.:\textit{ Directionality principle in thermodynamics and evolution,} Proc. Natl. Acad. Sci. \textbf{94} (1997), 3491--3498.	
	\bibitem{D2} Demetrius, L.: \textit{Boltzmann, Darwin and Directionality theory,} Physics Reports \textbf{ 530 }(2013), 1--85.
	\bibitem{D3}Demetrius, L.: \textit{Demographic parameters and natural selection,} Proc. Natl. Acad. Sci. \textbf{71} (1974), 4645--4647.
 	\bibitem{DG1}Demetrius, L., Gundlach, V.: \textit{Directionality Theory and the Entropic Principle of Natural Selection}, Entropy \textbf{16}, (2014), 5428--5522.
	\bibitem{E} Eigen, M.: \emph{Steps towards life: A perspective on evolution}, Oxford University Press, 1996.
	\bibitem{Fi}Fisher, R.A.: \emph{The genetical theory of natural selection}, The Clarendon Press, 1930.
	\bibitem{Ki} Kitchens, B.: Symbolic Dynamics: One--sided, two sided and countable state Markov shifts, Springer Verlag, Berlin Heidelberg 1998.
	\bibitem{Leh}Lehninger, A.: Bioenergetics, W.A. Benjamin, 1965.
	\bibitem{LL}Levins, L. and Lewontin, R.: The Dialectical Biologist, 1985
	\bibitem{MU} Mauldin, D. and M. Urbanski: Graph directed Markow Systems: Geometry and Dynamics of Limit Sets, Cambrigde: Cambridge University Press, 2003.
	\bibitem{PNB}Prigogine, I., Nicolis, G. and Babloyantz, A.: \emph{The thermodynamics of evolution}, Physics Today {\bf 25} (1972), 11,23. 
	\bibitem{R} Ruelle, D.: Thermodynamic Formalism, Cambridge University Press, Cambridge, 2004.
	\bibitem{SO}Sarig, O.: Thermodynamic formalism for countable Markov shifts, Proc. of Symposia in Pure Math. {\bf 89} (2015), 81--117.
	\bibitem{Sch}Schr\"odinger, E.:  \emph{What is life?}, Cambridge University Press, 1944.
	\bibitem{St}Stearns, S.: \emph{The Evolution of Life Histories,} Chapman \& Hall, 1992.
	\bibitem{Sp}Spiegelman, S : \emph{An approach to the experimental analysis of precellular evolution,} Quarterly Reviews of Biophysics {\bf 4} (1971), 213--253.
	\bibitem{W} Walters, P: \emph{An introduction to ergodic theory}, Graduate Texts in Mathematics 79, Springer, 1981.
	\bibitem{Wi} Wicken, J. S.: \emph{Evolution, Thermodynamics and Information}, Oxford University Press, 1987.
	\bibitem{ZD}Ziehe, M. and Demetrius L., \emph{Directionality theory: an empirical study of an entropic principle in life-history evolution}, Proc. Biol. Sci. {\bf 272} (2005), 1185--1198.

 



\end{thebibliography}
\end{document}